\documentclass{skaox2006}
\usepackage{graphicx}
\begin{document}
\title{Goonhilly Sparklers}   
\author{Aris Karastergiou\inst{1} \and Mark Walker\inst{2}}   
\institute{Oxford Astrophysics, Keble Road, Oxford, OX1 3RH, UK
  \and
  Manly Astrophysics, 3/22 Cliff Street, Manly 2095, Australia}    

\abstract{
Flux monitoring of compact radio quasars has revealed dramatic radio-wave lensing events
which challenge our understanding of the interstellar medium. However, the data on these
events remain very sparse. Here we consider how the Goonhilly radio astronomical facility
can make an impact on this problem by dedicating one or more dishes to flux monitoring for
a period of one year. Such an experiment would be able to identify $\sim6$ new events
and study them in detail.
}
\maketitle

\section{Introduction}

Extreme Scattering Events (ESEs) are a type of transience in which the
flux variations are not intrinsic to the source but are caused by
radio-wave refraction in ionised gas along the line-of-sight (Fiedler
et al 1987; Romani, Blandford and Cordes 1987). To generate these
events the ionised gas must have a pressure which is a thousand times
higher than the general Interstellar Medium (ISM), in a region of
dimensions $\sim1\;$AU. Thus the ESE phenomenon poses a serious
challenge to our understanding of the most basic physics of the ISM.

Because of the difficulty in understanding ESEs within a conventional
picture of the ISM, we have previously proposed that they are caused
by baryonic dark matter (Walker 2007; Walker and Wardle 1998): cold,
dense, AU-sized molecular clouds ramming their way through the ISM at
high speeds.

ESEs merit further study. Unfortunately there has been no substantial
new dataset since the original work of Fiedler et al (1987,1994) using
the Green Bank Interferometer. Within the next five years that
situation will change drastically as SKA-pathfinder instruments are
brought into service. In particular the Variables And Slow Transients
(VAST) project, which utilises the Australian Square Kilometre Array
Pathfinder (ASKAP: Johnston, Feain and Gupta 2009), will survey a
large fraction of the sky on a daily basis. But ASKAP only operates
efficiently up to 1.5 GHz, whereas the data we have on ESEs are at 2.7
GHz and above, making it difficult to plan for ESE science with
VAST. With Goonhilly we will change that, by discovering a number of
new events using 5 GHz data and then studying them at lower
frequencies. In the process we will gain some powerful new insights
into the physics of ESEs.

\begin{figure*}[!ht]
\begin{center}
\includegraphics[width = 12cm, clip = true, bb = 70 350 550 700]{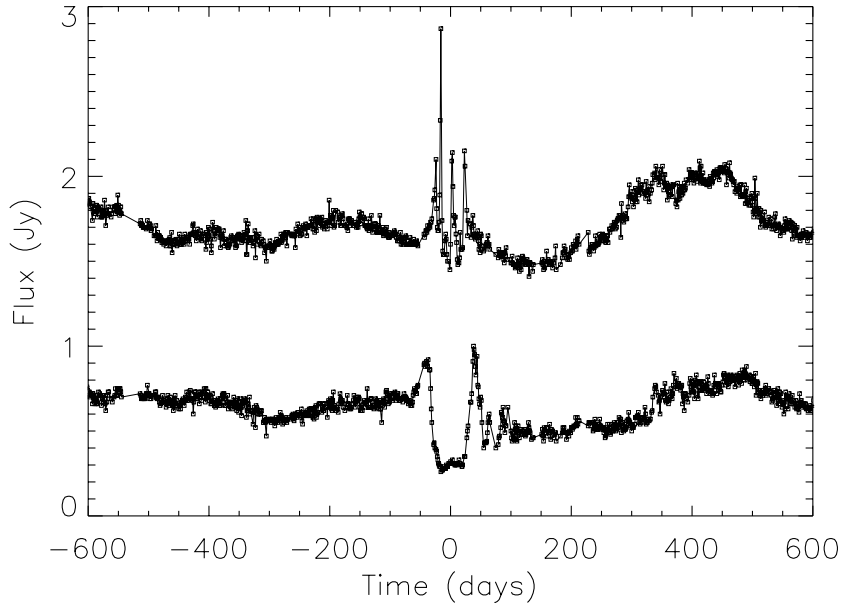}
\end{center}
\vskip -1cm
\caption{The best example of an Extreme Scattering Event: that
  detected in the source Q0954+658, at 8.1~GHz (top curve) and 2.7~GHz
  (lower curve), from Fiedler et al (1987).  An offset of $+1$~Jy has
  been added to the high-frequency data for clarity of presentation.
  These data taken from {\tt http://ese.nrl.navy.mil/.\/}
}\label{fig1}
\end{figure*}

\section{The Goonhilly facilities}
The Goonhilly Earth Station, in Cornwall, UK, was formerly a
telecommunications facility. On site are three 30m dishes and some
smaller (15m) antennas.  The Consortium of Universities for Goonhilly
Astronomy plans to instrument two of the 30m dishes for radio
astronomy, potentially including various frequency bands within the 1
to 10 GHz range.

In this paper we outline the possibilities for ESE science which may
be opened up by operating Goonhilly as an astronomical facility.  We
consider two hypothetical instruments:\hfill\break ${\bf C_1}$: a
single 30m dish equipped with a 5~GHz receiver having 1~GHz
bandwidth\hfill\break ${\bf C_2}$: a pair of 15m dishes operating as
an interferometer and receiving across the full 4 to 8~GHz band
simultaneously.\hfill\break

\section{ESE science with the ${\bf C_1}$ system}
To find ESEs we need to monitor the fluxes of a large number of
compact radio sources. The combined probability of the two most
striking ESEs (those in Q0954+658 and Q1749+096) is
$\sim5\times10^{-4}$, so that in a sample of 2,000 extragalactic
sources there will typically be one ESE in progress at any given
moment. And the event durations are $\sim2\,$months, so the event rate
in a sample of 2,000 sources is $\sim6\;{\rm year^{-1}}$.

For daily, year-round monitoring we need to choose sources which are
away from the ecliptic. Restricting ourselves to the region with
ecliptic latitude greater than $30^\circ$ means that we have $\pi$
steradians available, in principle.  But the reality of an alt-az
mount is that it can take a long time to slew between sources which
are North of the zenith, and those to the South, whereas we need to
minimise overheads associated with slewing. The latitude of Goonhilly
is approximately $50^\circ$ and we therefore restrict ourselves to the
1.5~sr between declination $50^\circ$ and the North Celestial Pole.

Our program sources must be compact, in order that they can be
significantly magnified, so only about one in 5 radio sources will be
suitable for our purposes. (Compact sources can be selected on the
basis of their radio spectra, which should be inverted or at least
flat.) To assemble a sample of 2,000 compact sources we must therefore
range down to $S_{min}\simeq35\,$mJy, where the areal density of all
radio sources is approximately ${\rm 6,\!000\,sr^{-1}}$ at 5~GHz (Wall
1994).

Assuming that the receiver is a clone of the C-BASS system the flux
noise for ${\bf C_1}$ should be approximately $4\,{\rm
  mJy\sqrt{s}}$. To detect magnification changes of order 10\% with
high confidence requires a signal-to-noise ratio of at least 30, which
would be achieved in about 12 seconds for a 35~mJy source.  (The
brightest confusing source in the beam will typically be 0.8~mJy, and
this source will normally be steady and will not be detrimental to our
study.) Thus the total required on-source time is less than 7 hours
per epoch for a sample of 2,000 targets. To this we must add the time
required for slewing and settling.

The typical angular separation between targets is less than
$2^\circ$. That is not a long slew and it seems reasonable to expect
that this can be achieved in under 30 seconds. If so, a sample of
2,000 sources can be monitored on a daily basis using ${\bf
  C_1}$. Daily sampling is desirable, even though the events last for
many weeks, in order to clearly distinguish ESEs from other forms of
variability and, especially, to do so in real time.

Thus with ${\bf C_1}$ operating for one year we can expect to detect 6
ESEs: a significant increase on the current sample of two. But the
greatest benefit is not so much the increase in event numbers as the
opportunity to identify those events in real time and thus to
characterise each one in detail. As well as detailed radio studies
(e.g. long baseline imaging), which constrain the ionised gas profiles
of the lenses, we aim to test for the presence of underlying neutral
gas, e.g. by the associated UV extinction.

\section{ESE science with the ${\bf C_2}$ system}
The ${\bf C_2}$ system has four times the bandwidth but only half the
collecting area of ${\bf C_1}$, so assuming the same system
temperature implies that it would take roughly the same integration
time to reach the same flux limit (assuming a flat spectrum source).

Confusion noise is potentially larger for ${\bf C_2}$, with the
brightest confusing source in the beam being typically about
3.5~mJy. However, if each target is observed at the same hour angle
(near transit, say), at every epoch then confusing sources make a
constant contribution to the measured visibilities and are not
detrimental to our study.

On the basis of experience with C-BASS it is anticipated (Mike Jones,
personal communication) that ${\bf C_1}$ should be photometrically
stable (i.e. to within the thermal noise) over periods up to about 100
seconds. It will therefore be necessary to intersperse our target
sources with some (steep-spectrum, non-compact so non-variable)
sources for calibration. Very few such sources would be needed for
${\bf C_2}$, because of the inherent stability of an interferometer,
and ${\bf C_2}$ could thus tackle a somewhat larger target sample than
${\bf C_1}$.

But the greatest advantage of ${\bf C_2}$ lies in the broad bandwidth
which it covers. Over an octave in radio frequency the refraction
angles introduced by the lens change by a factor of four. In turn this
means that the monitoring data themselves would yield tight
constraints on the electron column-density of the lenses, even in the
absence of any detailed real-time follow-up on other telescopes. Thus
the ${\bf C_2}$ system is superior to ${\bf C_1}$.

Lastly we note that for a long-term experiment, running for many years
and discovering large numbers of events, it will not be possible to
use facility-class instruments, like the VLA, to study each event in
detail.  It is here that the ${\bf C_2}$ system really comes into its
own because as a stand-alone experiment it yields much more powerful
constraints on the lenses than ${\bf C_1}$.
 


\end{document}